# Anomaly detection in wide area network mesh using two machine learning anomaly detection algorithms

James Zhang, Ilija Vukotic, Robert Gardner
University of Chicago

**Abstract**

Anomaly detection is the practice of identifying items or events that do not conform to an expected behavior or do not correlate with other items in a dataset. It has previously been applied to areas such as intrusion detection, system health monitoring, and fraud detection in credit card transactions. In this paper, we describe a new method for detecting anomalous behavior over network performance data, gathered by perfSONAR, using two machine learning algorithms: Boosted Decision Trees (BDT) and Simple Feedforward Neural Network. The effectiveness of each algorithm was evaluated and compared. Both have shown sufficient performance and sensitivity.

## 1. Introduction

A network anomaly is a deviation from the normal operation of the network, which is learned through observation and is signified by decreased network performance. In this paper, we are interested in four network performance measurements: throughput, packet loss rate, one-way delay (OWD), and traceroute. Throughput measures the amount of data that can be transferred over a time interval. The interval was chosen to be 25 seconds and is deemed not too long to cause undue stress on a link yet not too short to have unreliable measurements. Packet loss is the percentage of lost packets over the total transferred packets. To reach a sensitivity of $10^{-5}$ we measure it at 10 Hz and average result in one-minute bins. One-way delay measures delay (in ms) separately for each direction of a path. Traceroute is the path and transition time between the source and destination.

Anomalies can last anywhere from one hour to multiple days (if not noticed) and can be caused by a multitude of factors. Possibilities include: full connectivity disruption, in which all packets are lost; a device on the path is close to saturation, signified by an increase in one-way delay as packets spend more time in the device's buffers; a device on the path is saturated, where packets are lost as a device's buffers overflow, signified by increase in one-way delay; routing changes leading to asymmetrical paths which takes more time to reorder packets, signified by large variance in one-way delays; or dirty fibers or other problems with optics, signified by an increase in packet loss rate.

We collect our data from perfSONAR [1] servers. perfSONAR is a network measurement toolkit that monitors and stores network performance data between pairs of endpoints ("links"). While PerfSONAR is installed on thousands of servers, we are especially interested in ones that are part of the WLCG (Worldwide LHC Computing Grid) and OSG (Open Science Grid) meshes.

The network mesh size is extremely large. With an ever-increasing amount of links and issues in network performance, it becomes increasingly difficult to identify where, when, and why these issues arise and whether or not they are significant enough to ignore. Furthermore, due to the high variance and quantity of data collected, it becomes difficult to analyze and develop models of normal network behavior and of anomalies. Machine learning algorithms are thus favorable as they can learn what is normal behavior and subsequently what is anomalous behavior and adapt to changes in the structure of normal data.

Ideally an optimal anomaly detection method would satisfy the following criteria: Have the capability to naturally combine disparate data features and even data sources, e.g. all the links to or from a site, Packet loss, OWDs, paths, throughputs, FTS measurements, etc.; give information on what features (combinations of features) are causing the anomaly, and alert appropriate persons responsible for solving the issue; have tunable sensitivity, as we are not interested in short duration (order of 1 hour) flukes as no action can be taken at time scales shorter than that; and perform at a practical level.

*Packet loss*

We are especially interested in packet loss due to its extraordinary influence on throughput. As seen in Figure 1, even a 0.02% change in packet loss can cause a $10^3$ magnitude change in throughput.

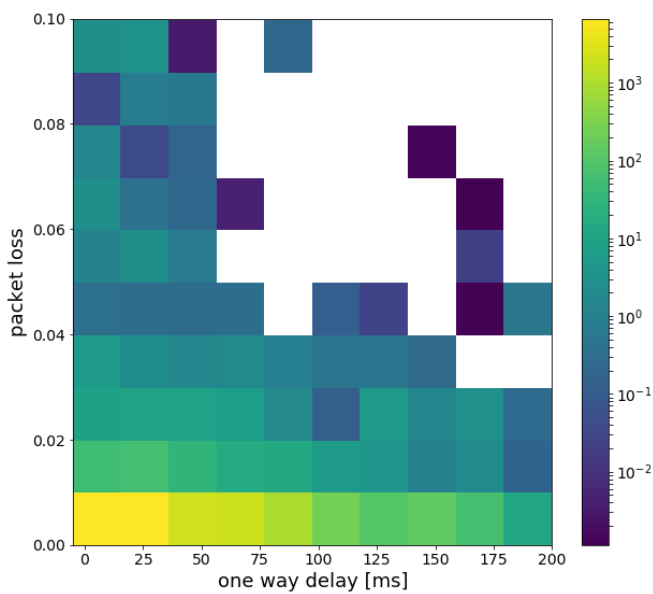

**Figure 1.** The relation of packet loss (%) to throughput.

*Related Work*

Machine learning techniques have been widely used in anomaly detection [2]. Su presented a density method of using the k-nearest neighbors algorithm (kNN) to detect denial-of-service attacks [3]. Sakurada et al. published a reconstruction method of using auto autoencoders to detect anomalies on both artificial data generated from Lorenz system and real data from a spacecraft's telemetry data [4]. Rajasegarar et al. described a boundary method of using one-class support vector machines for anomaly detection in sensor networks [5]. Recently, Catmore proposed a split-sample classification method and other ideas for anomaly detection [6]. This paper presents two new methods based on split-sample classification and reconstruction [6].

**2. Datasets**

We use two different datasets to test the functionality of the new machine learning algorithms as applied to network anomaly detection. Figure 2 shows an example of simulated data.

*Simulated Dataset*

A simulation of the actual data was generated to test the functionality of our methods. We defined six time series (features) spanning a seven-day period from 08-01-2017 00:00:00 to 08-07-2017 23:59:59. Data for each time series was assigned a value between 0 and 1 and was generated for each second. Normal data was generated for each time series by generating a random value less than 0.5 and then generating random values with a normal distribution between 0.00625 and 0.05 around that number. Normal data was flagged as 0. Table 1 is an example of such generated data. Anomalous data was 2 or 5 standard deviations from the normal noise distribution. Anomalous data was generated six times throughout the entire time period and had a maximum duration of 4 hours. Table 2 is an example of generated anomalous data. Figure 2 is an example of simulated data.

|  | link 0 | link 1 | link 2 | link 3 | link 4 | link 5 | flag |
|---|---|---|---|---|---|---|---|
| **2017-08-01 00:00:00** | 0.234309 | 0.098391 | 0.546828 | 0.290921 | 0.403015 | 0.245851 | 0 |
| **2017-08-01 00:00:01** | 0.230034 | 0.134577 | 0.548541 | 0.288414 | 0.399936 | 0.238371 | 0 |
| **2017-08-01 00:00:02** | 0.225711 | 0.127967 | 0.539710 | 0.287345 | 0.393710 | 0.243551 | 0 |
| **2017-08-01 00:00:03** | 0.231239 | 0.127110 | 0.543637 | 0.283620 | 0.401527 | 0.236844 | 0 |
| **2017-08-01 00:00:04** | 0.231674 | 0.141980 | 0.546474 | 0.290413 | 0.396456 | 0.235442 | 0 |

**Table 1.** The first five seconds of generated normal data.

| Time | Affected |
|---|---|
| 2017-08-03 07:36:42 2017-08-03 07:58:06 | [2, 5] |
| 2017-08-01 06:23:52 2017-08-01 07:06:09 | [2, 0, 1, 4] |
| 2017-08-05 18:30:38 2017-08-05 19:24:01 | [1, 3, 2, 4, 5] |
| 2017-08-02 11:27:58 2017-08-02 12:21:16 | [5, 2, 3] |
| 2017-08-05 07:20:14 2017-08-05 10:35:35 | [2, 1, 4, 0, 5] |
| 2017-08-03 19:20:06 2017-08-03 20:17:46 | [3, 1] |

**Table 2.** The time period and features affected for each anomaly generated by the simulated data. Feature numbers were listed in descending order of significance towards the data.

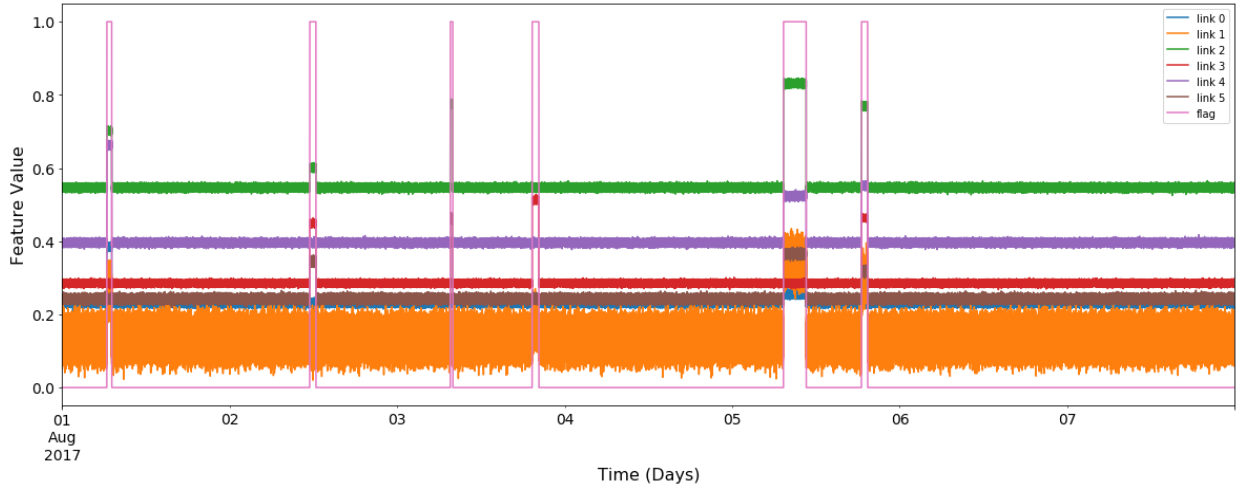

**Figure 2.** This figure is an example of simulated data. Each link, all assigned a specific color, is a different feature of the dataset and the pink columns signify anomalies. The vertical axis is the value of the data point for each feature and has arbitrary units.

*Real-World Datasets*

The data collected between various links measuring their one-way delay, throughput, and packet loss was used to test the functionality of our methods to see if they work on a practical level. Figures 3a and 3b show examples of real-world data that would be analyzed.

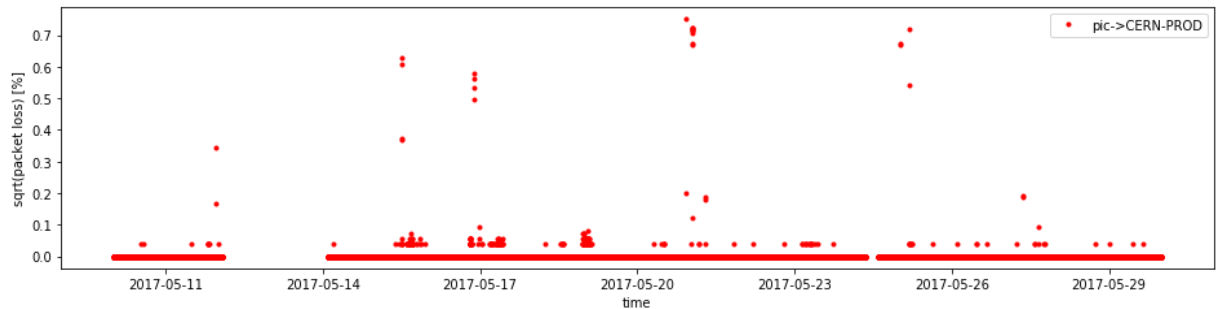

**Figure 3a.** This graph shows packet loss measured over a 20 day period between 2017-05-10 and 2017-05-30 for one link where the source is PIC and the destination is CERN. CERN is a Tier-0 site in Geneva, Switzerland, PIC is a Spanish Tier-1 center.

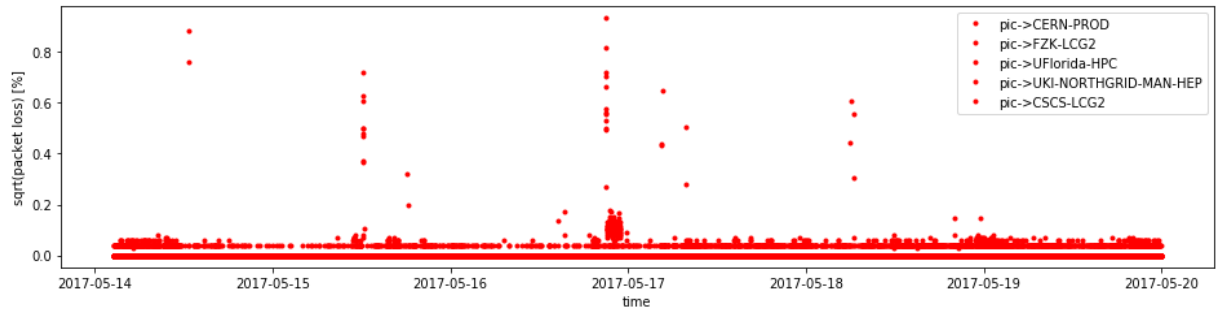

**Figure 3b.** This graph shows the square root of packet loss (%) over a six-day period between 05-14-2017 and 05-20-2017 for 5 sites: CERN-PROD, FZK-LCG2, UFlorida-HPC, UKI-NORTHGRID-MAN-HEP, CSCS-LCG2.

**3. New Anomaly Detection Method**

There are numerous algorithms for anomaly detection in time series. Among more general methods used are eg. ARIMA, SVM, and more specific ones eg. Bayesian inspired approaches [7]. We have no annotated historical data so that limits us to unsupervised methods. The amount of data that has to be continuously processed excludes several otherwise promising approaches. Similar to the split-sample classification method [6], in the following we compare time-dependent features in the period under examination (subject period) and in the directly preceding period (referent period). We do this by trying to train a Boosted Decision Tree (BDT)[8] or a Neural Network (NN) to correctly assign unlabeled samples to subject or referent periods. We expect that any significant difference in data between the two periods will be exploited by the BDT/NN. It follows that the accuracy of the classification will be proportional to the dissimilarity of the samples. We selected the referent periods of 24 hours as significantly long to not be sensitive to short duration incidents but short enough to capture long-term changes in link performance. Subject period of one hour is roughly the minimal period that one could expect a human intervention to happen in case alert was received. For both approaches, we flag the reference data with zero and the subject data with one. Further, 70% of the data from both the reference and subject periods are combined and used to train the machine learning models. Training effectiveness was then tested on the remaining 30% of the data.

*A. Boosted Decision Trees*

A decision tree is a rule-based learning method [2] that creates a classification model, which predicts the value of a target variable by learning simple decision rules inferred from the data features. A decision tree of depth 1 is known as a decision stump.

Decision trees make a split based on the Gini impurity, a measure of how often a randomly chosen element from the data set would be incorrectly labeled if it were labeled randomly. A higher Gini impurity suggests a less pure split.

For a set of items with $A$ classes, where $i \in \{1, 2, \ldots, A\}$. Let $p_i$ be the fraction of items labeled as class $i$ in the set. The Gini impurity can be calculated as follows [9]:

$$I_g(p) = 1 - \sum_{i=1}^{A} p_i^2$$

As examples of decision trees and stumps, Figures 4a and 4b show decision trees of different depths.

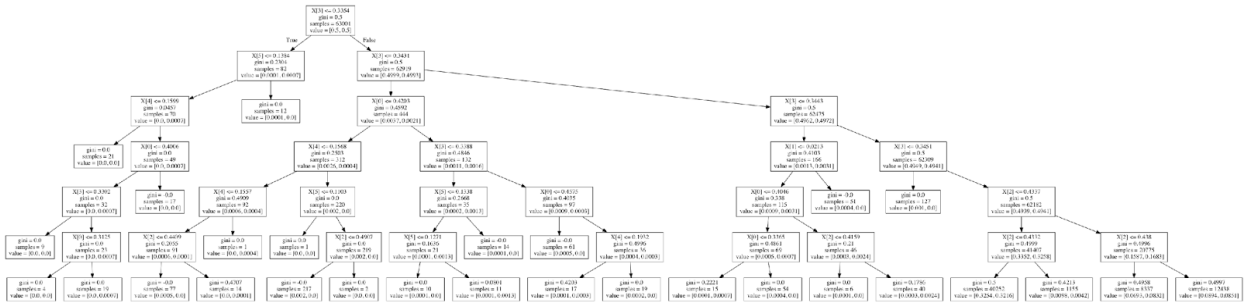

**Figure 4a.** This diagram is an example of a decision tree of depth 6.

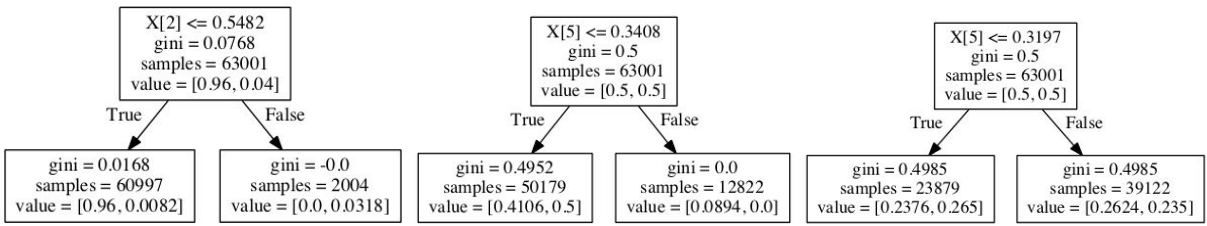

**Figure 4b.** This diagram is an example of three different decision stumps or decision trees of depth 1. Taking the first stump as an example, the tree splits the data on feature 2 with a Gini impurity of 0.0768 taking 63001 pieces of sample data. It was able to separate the data whereby the right column has a Gini impurity of 0.0, signifying that it was perfectly categorized.

We apply AdaBoost [10], a boosted decision tree algorithm, to our datasets. Boosting is a family of machine learning algorithms that start from weak classifiers, i.e. classifiers that label examples slightly better than random guessing, and return a weighted vote of all of them, the result of which is much more accurate at labeling.

Boosted decision trees train a decision tree by dividing data on one of the features. Then, it validates the decision tree against the training data to find misclassified data values. Each value is assigned a weight, determining its importance. The weight of misclassified values is increased and the weight of correctly classified values is decreased, then a new tree is built. Afterwards, it uses the original tree and the newly formed tree and tests against the training data again to find misclassified values. Another tree is formed, and this process is repeated for however many weak classifiers, i.e. estimators, desired. Each tree will produce a weighted vote between 0 or 1 on whether or not something is an anomaly. The majority vote is used to determine the classification (e.g., anomaly or not) of datasets.

*B. Simple feedforward neural network*

With recent advances in both hardware performance and availability (GPUs), and software stack (Keras, Tensorflow), it became possible to relatively quickly train neural networks with a large number of trainable parameters. We chose to start with a neural network consisting of one input layer with as many ReLU activation neurons as the number of time series data under investigation, one hidden layer with twice as many ReLU neurons and one sigmoid activated output neuron. We selected this topology as a surely sufficient to capture any effects in training data. In a future work, we will investigate the performance of a simpler, one layer network as this would simplify finding which time series contributed most to a period being flagged as anomalous. In order to increase performance, will also optimize learning rates, try different neuron activations, optimizers, etc.

Training and testing data were prepared in the same way as for the BDT model, except that training data gets shuffled after each epoch. We train for 60 epochs in batches of 256 samples using Adam optimizer [11] and as a loss function use binary cross-entropy. Training for one period takes around 20 seconds on a Tesla 20K NVidia GPU and roughly three times less on NVidia GTX 1080Ti. How well the trained network performed on the test samples is given by the binary accuracy. An anomaly is flagged based on how likely is the accuracy of that magnitude likely to happen by pure chance.

## 4. Experimental Results

*A. Boosted Decision Trees*

In this section, we study the effectiveness of BDT in detecting unusual events in network performance data. We primarily analyze packet loss and one-way delay. 50 estimators were used on the testing data. An AUC score was determined, and if the score was above a certain threshold then the result was determined to be an anomaly.

*A.1. Decision Stumps vs. Decision Trees*

We tested the effectiveness of both boosted decision trees utilizing trees of depth 6 vs. that of boosted decision trees utilizing decision stumps on our simulated data. It can be seen in Figures 5a and 5b that using decision stumps versus using decision trees produced roughly equal results. Both methods were able to identify anomalies of one hour in length. It is interesting to note that both methods were not able to detect the first anomaly. This is most likely because those anomalies occurred before the first day (our method uses a day's worth of data as training data), and as such did not have enough training data previously to identify them.

Though the two results were very comparable, the decision stumps took on average 4 seconds per hour worth of data to run, whereas decision trees took 15 seconds per hour worth of data, nearly four times longer. We determined that decision stumps applied to BDT were a more practical and efficient method for anomaly detection and were thus used for the remainder of the experiment.

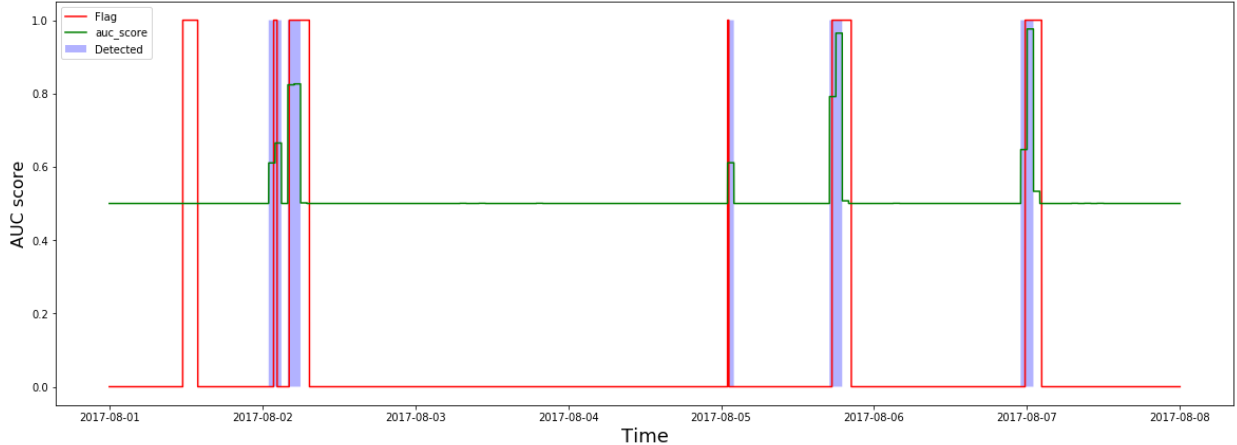

**Figure 5a.** This graph shows the results for BDT using decision stumps. The anomaly detection was performed on simulated data of a period of 7 days. Red lines indicate where anomalous data was generated, blue shading indicates where the algorithm predicted an anomaly was, and the green is the AUC score for each hour.

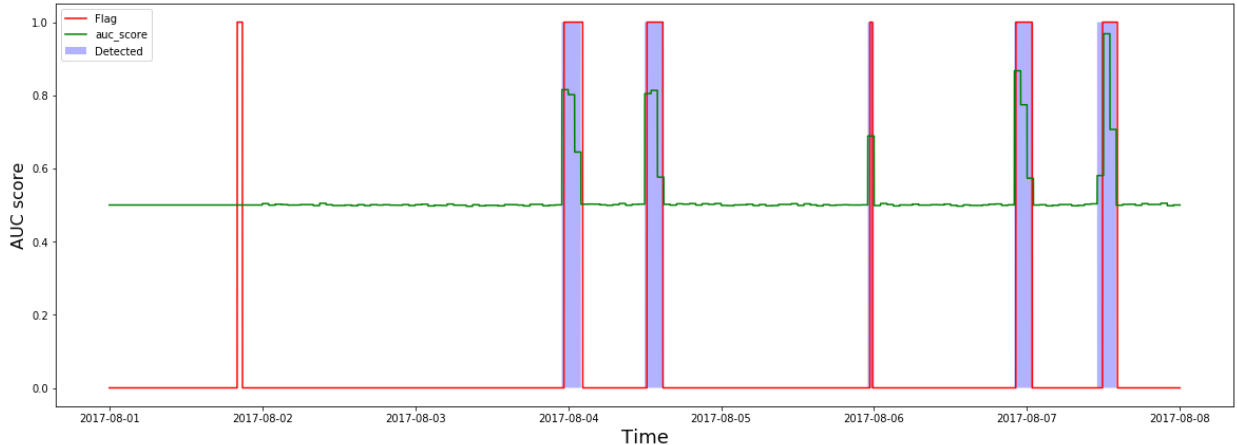

**Figure 5b.** This graph shows the results for BDT using decision trees of depth 6.

*A.2. Anomaly Duration vs. Degree of Anomalous Behavior*

We wanted to know which had a greater effect on whether an interval was determined to be anomalous: the magnitude of the offset between an anomaly from normal data, the duration of anomalous behavior, or the number of time series affected. Six anomalies were generated on the same set of normal data. Figure 6 shows a visual representation of said generated anomalies. The start of each anomaly was separated by 24 hours. We fluctuated the duration of anomalous behavior, the number of features affected, and the anomaly offset for each one. The degree of anomalous behavior encompasses both the number of features affected and the anomaly offset.

Data considered to have affected many features affected three times as many as data that was not (3 features vs 1). Data with a long duration of anomalous behavior was three times longer than data (3 hours vs 1). We generated offsets (anomalies) of two amplitudes: small - $2\sigma$ and large - $5\sigma$ shift. Normal data had an accuracy of roughly 0.50. The threshold for anomalous behavior was determined as any AUC score above 0.55. From the results shown in Table 3 and Figure 6, we see that the change in the anomaly offset had the most significant effect on whether an anomaly was determined to be so.

Holding the number of features affected and the duration of anomalous behavior constant, an increase in anomaly offset increased the AUC score by 0.379, a roughly 63.91% increase. The number of features affected had a less significant effect. Holding the anomaly offset and anomaly duration constant, an increase in the number of features affected increased the AUC score by 0.267, a roughly 45.03% increase. The duration of anomalous behavior had the smallest effect on determining an anomaly. Holding the other two variables constant, an increase in the duration of anomalous behavior decreased the AUC score by 0.068, a roughly 11.47% change. The results suggest that the extent to which an anomaly is considered as such is more dependent on the anomaly offset and the features affected, that is, the degree of anomalous behavior, more than the anomaly duration.

| Anomaly | Offset [$\sigma$] | Features Affected | Duration [h] | AUC score | | | |
|---|---|---|---|---|---|---|---|
| | | | | Hour Before | Hour 1 | Hour 2 | Hour 3 |
| 1 | 2 | 1 | 1 | 0.499 | **0.593** | | |
| 2 | 2 | 1 | 3 | 0.500 | **0.525** | **0.511** | 0.501 |
| 3 | 2 | 3 | 1 | 0.499 | **0.860** | | |
| 4 | 5 | 1 | 1 | 0.499 | **0.972** | | |
| 5 | 5 | 1 | 3 | 0.500 | **0.693** | **0.702** | 0.502 |
| 6 | 5 | 3 | 1 | 0.500 | **0.999** | | |

**Table 3.** This table shows the AUC scores of six different simulated anomalies all having some combination of anomaly duration, amplitude, and a number of features affected. Anomaly numbers come in the order of when the anomaly was generated. Results above our 0.55 cut level are shown in bold letters.

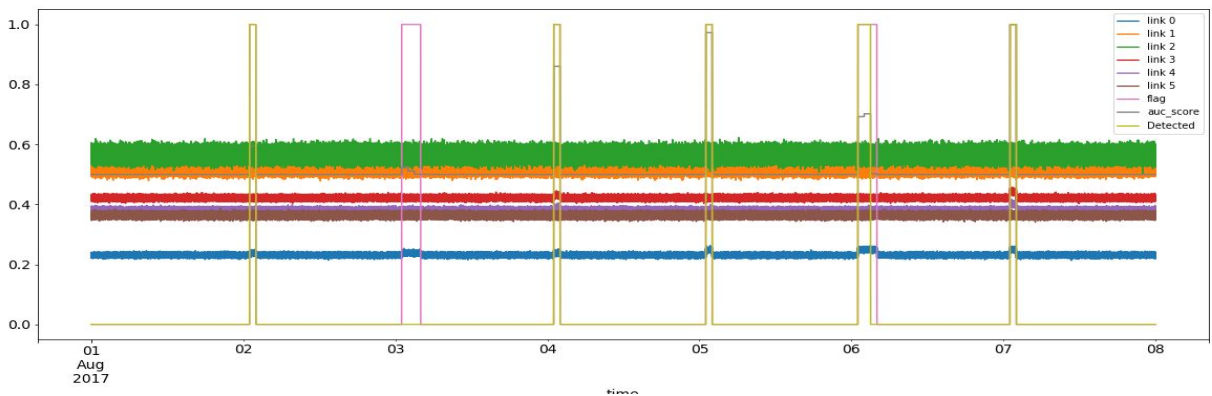

**Figure 6.** This figure shows 6 anomalies generated over a 7 day period. Each anomaly has some combination of anomaly duration, amplitude, and a number of features affected.

*A.3 Application on real data*

We tested the effectiveness of using a boosted decision tree onto both the packet loss and one-way delay data over a 4-day time span between two sites: PIC and CERN-PROD. The AUC threshold was set to 0.55. From Figure 7, it is evident that the AUC threshold is too high, as it detects anomalies far too often to be practical.

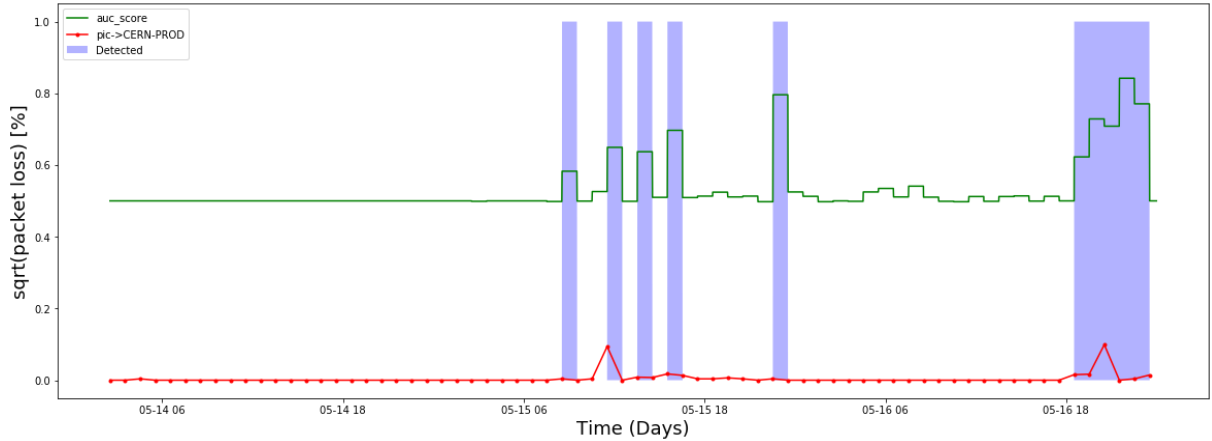

**Figure 7.** This graph shows the boosted decision tree as applied to one-way delay and packet loss data between PIC and CERN-PROD between . The light blue columns signify areas where the algorithm detected anomalous activity, using an AUC threshold of 0.55.

For each anomaly that was detected, we were able to generate both the feature importance of each time series and a ROC curve, thus allowing us to tell which time serie(s) caused the anomaly. Figure 8 shows an example of said importance and curve.

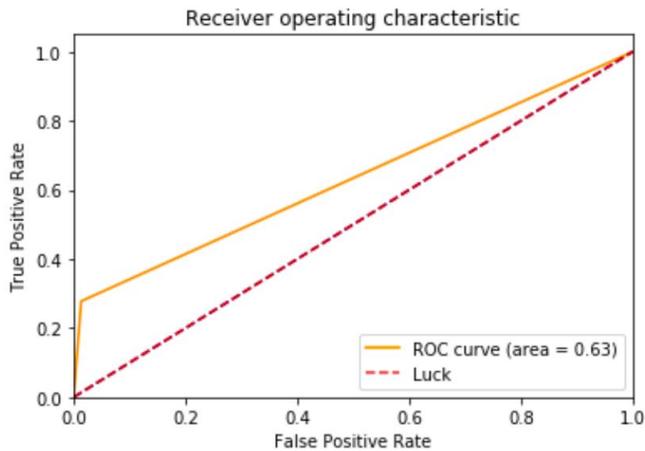

**Figure 8.** This figure shows the first anomaly detected, the relative importance of features, and its ROC curve. By looking at the feature importance, we see that feature 2 had the greatest influence on the AUC score, then feature 3, feature 1, and finally feature 4, which did not factor in at all in generating the anomaly.

Due to the high frequency at which anomalies were detected, the AUC threshold of 0.55 was determined to be impractical. After adjusting the AUC score to 0.8, the result as seen in Figure 9, became much more practical to be used.

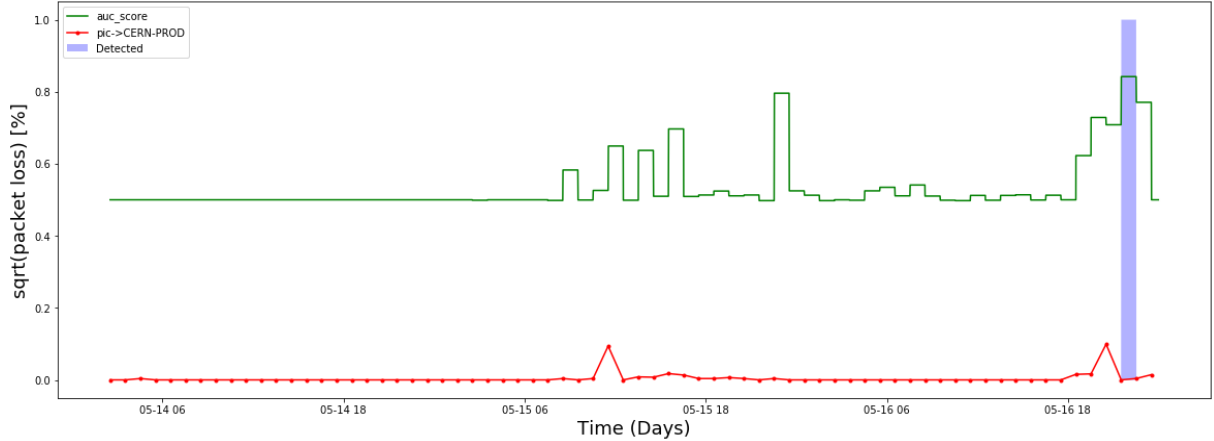

**Figure 9.** This graph shows the boosted decision tree as applied to one-way delay and packet loss data between PIC and CERN-PROD in a 30-day period. The AUC threshold was 0.80.

*B. Simple feedforward neural network*

In this preliminary study, we used two hidden layers with twice as many ReLU neurons as time series and a single sigmoid output neuron. This results in a network with 2521 trainable parameters.

The result of the test is a binary classification accuracy - defined as a ratio of correctly labeled and total number of samples. The distribution of accuracies that can be expected by chance depends only on the number of samples in referent and subject intervals and can be seen in Figure 10.

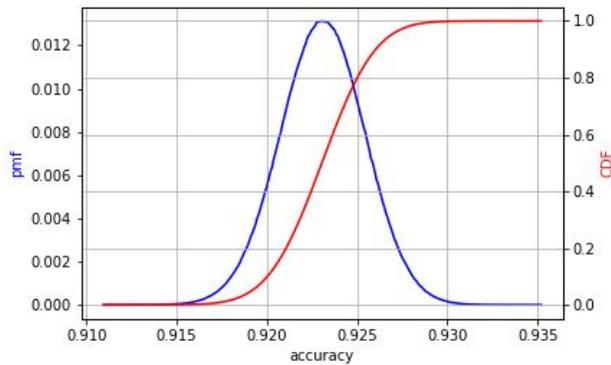

**Figure 10.** This figure shows the distribution of accuracies obtained by chance (binomial distribution) corresponding to 12h referent and 1h subject intervals with one sample per second (p=0.92308). Accuracy larger than 0.9285 can happen with <1% chance.

*B1. Applied to simulated data, impact of duration, magnitude of anomaly, and number of affected time series*

We used the simulated data described in *A2*. A 12h period prior to the 1h subject period was used as a reference. This choice gives an accuracy of 0.923 by pure chance, and we consider anomaly detected if the calculated accuracy has less than 1% chance of appearing randomly (0.9285). From results shown in Table 4 and Figure 11, we see that only the last three anomalies (with offsets of 5$\sigma$) have been detected, and only it the first hour of anomaly appearance. Accuracy levels during the first three anomalies and during the last two hours of the fourth anomaly were not simply under the threshold but actually exactly equal to pure chance accuracy. As it can be seen from Figure 12 loss and accuracy change in a stepwise manner and in case backpropagation optimization does not find any minima result will be equal to chance. This does not limit the applicability of the method as we are anyhow not interested in small effect anomalies affecting single time series, or repeated identification of the same anomaly.

| Anomaly | Offset [$\sigma$] | Features Affected | Duration [h] | Accuracy | | | |
|---|---|---|---|---|---|---|---|
| | | | | hour before | hour 1 | hour 2 | hour 3 |
| 1 | 2 | 1 | 1 | 0.923 | 0.923 | | |
| 2 | 2 | 1 | 3 | 0.923 | 0.923 | 0.923 | 0.923 |
| 3 | 2 | 3 | 1 | 0.923 | 0.923 | | |
| 4 | 5 | 1 | 1 | 0.923 | **0.942** | | |
| 5 | 5 | 1 | 3 | 0.923 | **0.991** | 0.923 | 0.923 |
| 6 | 5 | 3 | 1 | 0.923 | **0.999** | | |

**Table 4.** This table shows the accuracies of six different simulated anomalies all having some combination of anomaly duration, amplitude, and a number of features affected. Anomaly numbers come in the order of time the anomaly was generated. Results above our 0.928 cut level are shown in bold letters.

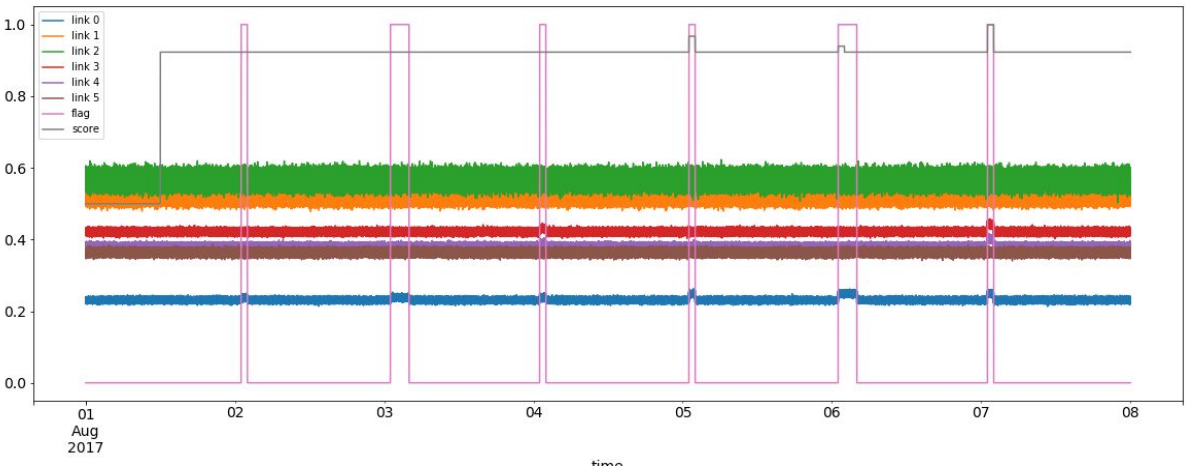

**Figure 11.** This figure shows 6 anomalies generated over a 7 day period. Anomalies are described in **Table 4**. Only the last three anomalies have been detected.

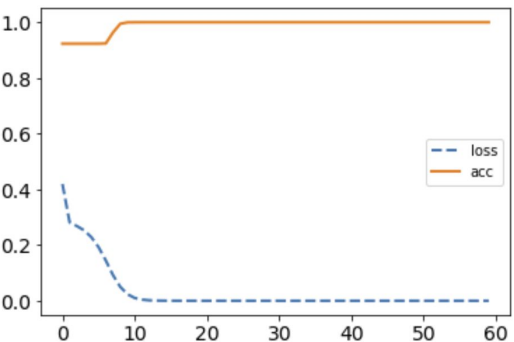

**Figure 12.** Loss and accuracy for each of 60 training epochs. Anomaly threshold was already reached at epoch 15. One possible optimization is early training termination as soon as the threshold has been reached.

*B3. Performance on actual data*

When using this method on actual data we use referent interval of 24 hours to average possible anomalies in referent data over longer periods. Data is measured packet loss between CERN and 20 other sites (from 73 that we have values measured for) covering a four day period. There is one data point per minute. Missing data was filled with zero values. Figure 13. shows results obtained by training the network for 100 epochs for each interval, using a batch size of 10 and having 70:30 split between training and testing data. Mean of the chance accuracy distribution is at 0.96 and chance accuracy of more than 0.97 has less than 1% probability.

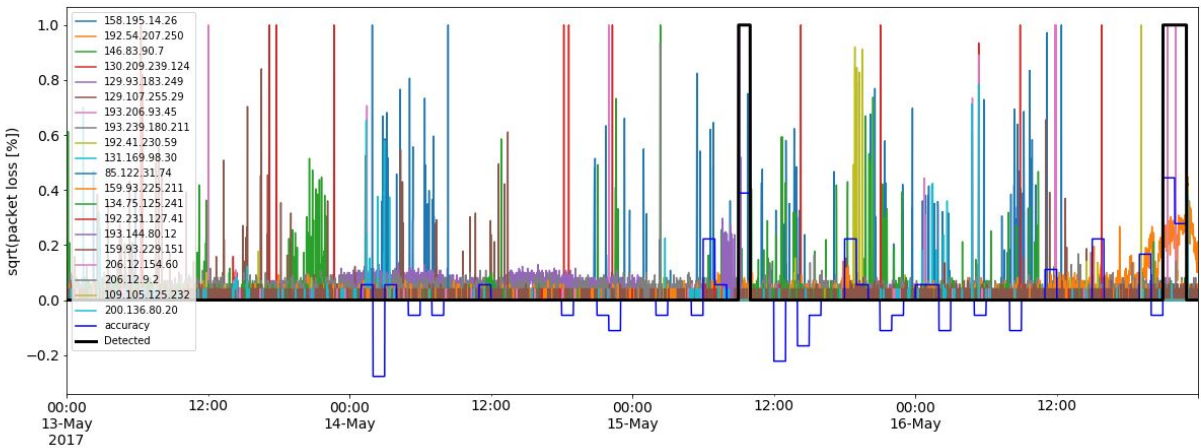

**Figure 13.** Packet loss between CERN and 20 other sites (here given as an IP address of their PerfSONAR node). Shown is the square root of the value to better visualize small values. The blue line shows binary classification accuracy improvement over the chance value. Thick black line ("Detected") marks intervals flagged as anomalous.

## 5. Conclusion

This paper presents two new methods of detecting network performance anomaly based on split-sample classification: AdaBoost and Simple feedforward neural network. Both methods are first tested on simulated datasets to check their sensitivity with respect to duration and amplitude of anomaly. The boosted decision tree method proved to be very fast (4 seconds evaluation per one hour of data tested) and detected all the simulated anomalies. An added benefit is that it directly returns ordered list of series according to their contribution to the anomaly being flagged. With appropriately selected AUC threshold it is possible to tune desired sensitivity/false positive level.

The simple neural network model used was not hyper-parameter optimized and the one network tried proved less sensitive to short and low amplitude changes. Given that we are looking for the most significant anomalies this is a good feature. While the evaluation is slower at 20 seconds per hour of data tested, it is still fast enough to be of practical use. A more significant issue is that it requires a GPU for processing. Since this is a three-layer network it is difficult to get information on the importance of different time series to the resulting decision. While results on the actual data are encouraging, before using it in a production environment, different network configurations should be tested (two layers, fewer neurons per layer, etc.) and hyper-parameter tuned.

# Appendix

All the codes and test data can be found in this repository:
https://github.com/ATLAS-Analytics/AnomalyDetection